\documentclass{amsart}

\usepackage{amssymb,amsfonts,latexsym,amsmath,amsthm,wrapfig,graphicx, hyperref, %showlabels
}
\input xy
\xyoption{all}

\numberwithin{equation}{section}
\newtheorem{thm}{Theorem}[section]

\newtheorem{remark}{Remark}
\newtheorem{definition}[thm]{Definition}

\begin{document}

\title[Projective connections and extremal domains for analytic content]{Projective connections and extremal domains for analytic content}

\date{August 2019}

\author[R. Teodorescu]{Razvan Teodorescu}
\email{razvan@usf.edu}
\address{4202 E. Fowler Ave., CMC342, Tampa, FL 33620}

\maketitle

\begin{abstract}
This note expands on the recent proof \cite{ABKT} that the extremal domains for analytic content in two dimensions can only be disks and annuli. This result's unexpected implication for theoretical physics is that, for extremal domains, the analytic content is a measure of non-commutativity of the (multiplicative) adjoint operators $T, T^{\dag}$, where $T^{\dag} = \bar z$, and therefore of the quantum deformation parameter (``Planck's constant"). The annular solution (which includes the disk as a special case) is, in fact, a continuous family of solutions, corresponding to all possible positive values of the deformation parameter, consistent with the physical requirement that conformal invariance in two dimensions forbids the existence of a special length scale. 
\end{abstract}

\section{Introduction}

The problem of determining extremal domains for analytic content has a long history, and a recent resolution (see \cite{ABKT}, and references therein, listed here as well)  of 
decades-old conjectures \cite{Kh84} has left only the infinite-connectivity case still open. Throughout these explorations, several connections to theoretical physics were identified, from Serrin's problem for laminar flows of Newtonian fluids through constant cross-section pipes  \cite{Se, K87, BeKh} to hydrodynamics of electrified droplets \cite{Ga,BeKh,KSV}, and extensions to multi-boundary conformal field theory \cite{RT2014}. 

We recall the general definition of analytic content in two dimensions: given a compact set $K \subset \mathbb{C}$, we denote by $R(K)$ the closure, in the set of continuous functions on $K$, $C(K)$ (with respect to the supremum norm $||\, ||_{\infty}$), of the set of rational functions with poles in the complement of $K$. The analytic content of $K$ is then given by solution to the best approximation, in $R(K)$, of the antiholomorphic coordinate, $\bar z$: 
\begin{definition}\label{ac}
    $$ \lambda(K):= \inf_{\varphi \in R(K)} \| \bar{z}-\varphi \|_{\infty}.$$
\end{definition}
Geometric constraints were found for the analytic content \cite{Al,Kh84}, relating it to the area and perimeter of $K$:%, much like the Weyl conjectures for the short-time expansion of heat kernel for the Laplace operator: 
\begin{equation}\label{geo}
    \frac{2Area(K)}{P(K)} \leq \lambda(K) \leq \sqrt{\frac{Area(K)}{\pi}},
\end{equation}
where $P(K)$ is the perimeter of $K$. These inequalities lead naturally to the definition of extremal domains for analytic content: 

\begin{definition}
Let $\Omega \subset \mathbb{C}$ be a bounded domain of finite connectivity $n \in \mathbb{N}$, and real analytic boundary $\Gamma = \cup_{k=1}^n \Gamma_k$. Then $\Omega$ is an extremal domain for analytic content if $\lambda(\overline{\Omega}) = 2 Area(\Omega)/P(\Gamma)$. 
\end{definition}

In  \cite{ABKT}, along with a proof of the conjecture \cite{Kh84} that for an extremal domain $n \le 2$, it was shown that one of the reformulations of this property translates into a conformal field theory (CFT) problem: find a domain $\Omega$ with boundary $\Gamma = \cup_{k=1}^n \Gamma_k$, endowed with a quadratic differential (classical (holomorphic) stress-energy tensor) $T_{zz} = \varphi'(z) dz^2$ and $n$ chiral fields $v_k(z)$, satisfying the projective connection null conditions \cite{Frenkel}
\begin{equation} \label{eq}
\left [ \frac{d^2 \,\,}{d z^2} + \frac{\varphi'(z)}{\lambda^2} \right ] v_k = 0, \,\, \forall \, z \in \Omega, \,\, k = 1, 2, \ldots, n.
\end{equation}
 
In this note, we prove that the only type of extremal domain (annulus, with the disk as a limiting case) corresponds to the unique geometry compatible with quantum deformations in 2D, and that the deformation parameter can be identified to the analytic content itself, and is given by the difference between the two annulus radii, $\lambda = R_1 - R_2 \ge 0$. The fact that the range of $\lambda$ is $\mathbb{R}_+$ is consistent with the existence of a classical limit for these quantum deformations, and which corresponds to the degenerate annulus of vanishing area (or empty interior).  This interpretation for the extremal domain problem (and solution) is provided in Section 2. In Section 3, we discuss the group of automorphisms of an extremal domain, and identify its generators to two generators of conformal transformations, related to the general Virasoro algebra. Using this interpretation, we further identify the constraints \eqref{eq} to the Kac-type conditions for minimal rational CFTs, thus proving that QFTs achieving the minimal value of the deformation parameter are conformal, and the group of automorphisms of the domain to the Lorentz subgroup generated by rotations and time reversal. Section 4 covers concluding remarks about a possible functional optimization formulation of the extremal domain problem.

\section{A quantum field-theoretic interpretation of the extremal domain problem}

As shown in \cite{ABKT}, if a domain $\Omega \subset \mathbb{C}$ of finite connectivity $n$ and real-analytic boundary $\Gamma$ consisting of $n$ disjoint components $\Gamma_{k}, \, k = 1, 2, \ldots, n$, is an extremal domain for analytic content, then the following are true:

\begin{itemize}
\item either $n = 2$ or $n = 1$, with the former corresponding to an annulus of boundary components $\Gamma_1 = \{ z | |z| =  R_1 \}$, $\Gamma_2 = \{ z | |z| =  R_2 \}$, and $\lambda = R_1 - R_2 \ge 0$, and the latter being the limit case $R_2 \to 0$ for the annular case;
\item  for both cases, the best approximation is given by the rational function 
$$
\varphi(z) = \frac{R_1R_2}{z},
$$
with the limiting case $\varphi(z) = 0$ for the disk case ($R_2 = 0$); 
\item in the general case $n =2, R_2 > 0$, there are two independent solutions $v_{1,2}(z)$ to the Ricatti-type equation 
$$
v_k''(z) = \frac{R_1R_2}{\lambda^2}\cdot \frac{v_k(z)}{z^2}, 
$$
namely
\begin{equation} \label{solutions}
v_1(z) = \left (\frac{z}{R_1} \right )^{\frac{R_1}{\lambda}}, \quad 
v_2(z) = \left (\frac{R_2}{z} \right )^{\frac{R_2}{\lambda}}, 
\end{equation}
so that $|v_k(z)|_{\Gamma_k} = 1$, and $v_1$ is bounded in $\{|z| \le R_1\}$, while $v_2$ is bounded in $\{|z| \ge R_2\}$. Obviously, in the disk limit case $n=1, R_2 = 0$, we have only one solution, $v_1(z) = z/R_1$. 
\end{itemize}

Let us consider now a Hilbert space $\mathcal{H}$ over $\Omega$, corresponding to a quantum deformation of a symplectic structure, and with deformation parameter $h \ge 0$. Identifying the adjoint multiplicative operators $T = (T^{\dag})^{\dag}$, $T^{\dag}(f) = \bar z f$ to elements from the algebra of bounded continuous operators $\mathcal{B}(\mathcal{H}) \subset L(\mathcal{H})$, we have a functional relation between the norm of the commutator $[T, T^{\dag}]$ and the deformation parameter $h$. While the exact functional relation will depend on the specific Hilbert space structure of $\mathcal{H}$ (functional model), and in particular on its induced norm on $\mathcal{B}(\mathcal{H})$, $||\,.\,||_{\mathcal{B}(\mathcal{H})}$, it is obvious that the optimization problems 
$$
\arg(\inf_{\varphi \in R(\overline{\Omega})} \| \bar{z}-\varphi \|_{\infty}), \,\, 
\arg(\inf_{\varphi \in R(\overline{\Omega})} \| T^{\dag}-\varphi(T) \|_{\mathcal{B}(\mathcal{H})})
$$
will share the same solution. In particular, this implies that the existence of a non-vanishing lower bound for the analytic content of $\Omega$ would translate into the existence of a strictly-positive lower bound for the possible values of the quantum deformation parameter $h$, and therefore an apparent violation of the classical correspondence principle. Therefore, we conclude that the only 2D geometry fully consistent with the axioms on quantization deformation corresponds to the family of extremal domains (annuli, $n=2$), parametrized by the analytic content $\lambda \ge 0$. 

Furthermore, we note that the chiral fields \eqref{solutions} are single-valued only if 
$$
\frac{R_1}{\lambda} \in \mathbb{N}, \quad 
\frac{R_2}{\lambda} \in \mathbb{N}, 
$$
which implies that $\frac{R_2}{R_1} = \frac{n-1}{n}, \,\, R_1 = n \lambda, \,\, R_2 = (n-1) \lambda,$ for some $n \in \mathbb{N}$, which means that, at fixed $\lambda \in \mathbb{R}_+$, there is only a countable family of acceptable domains, parametrized by $n \in \mathbb{N}$. In particular, this implies the area ``quantization" condition 
$Area(\Omega) = (2n - 1) \pi \lambda^2$. 

To complement these observations, we also note that, in the quantized case, the chiral fields $v_k(z)$ have the canonical boundary limits 
$$
v_1(\theta) = e^{in\theta}, \quad v_2(\theta) = e^{-i(n-1)\theta}, 
$$
for $z = e^{i \theta} \in \Gamma_{k}, \, k = 1, 2$, and are therefore eigenvectors of the generator of global rotations, $L_0 \in sl(2, \mathbb{C})$.  Up to overall constants, they are also related by the identities 
$$
v_1 = \left (\frac{d v_2}{dz}  \right )^{-1} = (R \circ L_{-1}) v_2, \quad 
v_2 = (R \circ L_{-1}) v_1 
$$
where we denote by $R$ the inversion transformation (up to overall constants) $R(z) = z^{-1}$, and $L_{-1} = - \frac{d}{dz}$ is one of the generators of $sl(2, \mathbb{C})$, as a subalgebra of the Virasoro algebra. This observation naturally leads to the more general question of determining the invariance group of an extremal domain, and then finding its irreducible representations. 

\section{The group of automorphisms of the extremal domain as QFT invariance group}

For the annulus $\Omega$ of radii $R_1, R_2$, the group of automorphisms $Aut(\Omega)$ consists of rotations $z \to z\cdot e^{i \alpha}$ and inversion with respect to the mid-circle $|z|^2 = R_1R_2$, or $z \to \frac{R_1R_2}{z}$. Since these are elements of $SL(2, \mathbb{C})$ leaving $\Omega$ (and therefore $\lambda(\Omega)$) unchanged, the projective connection \eqref{eq} will transform covariantly, implying that the chiral fields $v_k(z)$ must also transform covariantly under elements of $SL(2, \mathbb{C})$.  This leads to the more general question: given that $(\Omega, \lambda)$ is an extremal domain with analytic content $\lambda$, is it true that $(a\cdot \Omega, |a|\cdot \lambda)$ is also an extremal domain/analytic content pair, where $a \in \mathbb{C}^*$ is a non-zero complex number? 

The answer, as can be read off from Section~2, is affirmative, since indeed annuli (and their analytic contents) transform covariantly under global dilations and rotations. However, it should be noted that this approach would not have led easily to the proof of the conjecture that only annuli can be extremal domains, since the projective connections \eqref{eq} would not transform in a simple way under generic (complex) dilations. Nonetheless, having the solution in hand, we can reach some interesting conclusions from this invariance property. Adding the (obvious) invariance of the solution with respect to global translations, we conclude that the family of extremal domains is invariant under the M\"obius group, which implies that the corresponding quantum deformation problem must correspond to a conformal field theory. This fact has a number of interesting consequences. 

The first of these consequences is related to the topological interpretation of the annular solution. In the radial quantization of CFT, the spatial coordinate is compactified into the unit circle $S^1$, while the radial coordinate corresponds to time, leading to the annular geometry in which the circular boundary of lower radius $(R_2)$ corresponds to the initial time (and the in-states of the theory), while the second, concentric circular boundary ($R_1$) corresponds to the final time, and the out-states. Therefore, the annular solution features two primary fields, $v_2$ (of conformal weight $\Delta = \frac{R_2}{\lambda}$), and $L_{-1} v_2$ (of dimension $\Delta + 1$), whose time-reversal state is $v_1$.  

\begin{remark}
From this point of view, the connectivity constrain $n \le 2$ is immediate, as there is no other possible causal geometry in CFT, and a domain with $n \ge 3$ would imply the existence of ``temporal gaps" with no physical interpretation. 
\end{remark}

A second consequence is related to the physical interpretation of the group of automorphisms of $\Omega$. As noted above, it consists of rotations and inversions with respect to the mid-circle $|z|^2 = R_1R_2$. Physically, these correspond to the (Lie) subgroup of continuous spatial transformations that leave the time (radial) coordinate unchanged (global rotations), and of the discrete time reversal symmetry, which exchanges the in-states and out-states. 

\begin{remark}
We note again that the physical requirement for $Aut(\Omega)$ to contain a Lie subgroup (of spatial invariance) would have immediately imposed the condition $n < 3$ in the general extremal domain problem, due to the known result \cite{Heins} that for domains of connectivity $n \ge 3$, $Aut(\Omega)$ is a finite group. 
\end{remark}

A third observation one can make pertains to finite representations for the CFT associated with the extremal domain problem. As noted, the projective connection \eqref{eq} transforms covariantly under global dilations, while the solutions \eqref{solutions} are scaling functions, and can be identified to correlation functions for a critical system. Therefore, having correlation functions solve a second-order ODE leads to the identification of the equation \eqref{eq} to the simplest form of the Kac determinant constraint \cite{Kac, FeF}, at level 2, which will correspond to the existence of a ``null" vector (orthogonal to the irreducible representation) 
$$
|\chi\rangle = [L_{-2} + r(\Delta) L_{-1}^2 ] |\Delta \rangle, 
$$
where $r(\Delta)$ is a rational function of the highest weight exponent $\Delta$, which depends on the choice of central charge. Therefore, it is possible to associate to a family of extremal domains minimal rational CFTs (unitary, and of positive, rational critical exponents).

\section{Concluding remarks}
The brief discussion presented in this note can obviously be extended to more general dualities between extremal problems in approximation theory (of which the problem of extremal domains for analytic content is a prime example), and unitary representations of conformal invariance. One such extension can be obtained by expressing the extremal problem as an optimization problem in the Hamiltonian formalism of CFT. This would allow to generalize this class of problems by considering  optimization theory for other functional extensions.

\end{document}